\documentclass[a4paper,11pt]{article}
\usepackage{pos}

\title{A first study of strong isospin breaking effects in lattice QCD using truncated polynomials}
\ShortTitle{A first study of strong IB effects in LQCD using truncated polynomials}

\author*[a,b]{David Albandea}
\author[d]{Simon Kuberski}
\author[c]{Fernando P. Panadero}

\affiliation[a]{Helmholtz-Institut Mainz, Staudingerweg 18, 55128 Mainz, Germany}
\affiliation[b]{GSI Helmholtz Centre for Heavy Ion Research, 64291 Darmstadt, Germany}
\affiliation[c]{Instituto de Física Teórica UAM-CSIC, calle Nicolás Cabrera 13-15, 28049 Madrid, Spain}
\affiliation[d]{Theoretical Physics Department, CERN, 1211 Geneva 23, Switzerland}

\emailAdd{albanded@uni-mainz.de}

\abstract{Computing derivatives of observables with respect to parameters of the
theory is a powerful tool in lattice QCD, as it allows the study of physical
effects not directly accessible in the original Monte Carlo simulation.
Prominent examples of this include the impact of the up-down quark mass
difference and electromagnetic corrections. In this work, we present a new
approach based on automatic differentiation to evaluate such derivatives to
arbitrarily high orders, where particular emphasis will be placed on strong
isospin-breaking effects and on the propagation of derivatives through the
conjugate gradient algorithm in the computation of correlation functions.

\vspace*{0.5cm}
\begin{flushright}
    MITP-26-015\\
    IFT-UAM/CSIC-26-38 \\
    CERN-TH-2026-059
\end{flushright}
}

\FullConference{The 42nd International Symposium on Lattice Field Theory (LATTICE2025)\\
2-8 November 2025\\
Tata Institute of Fundamental Research, Mumbai, India\\}

\newcommand{\kaonlo}{\raisebox{-0.4\totalheight}{\includegraphics[scale=.4]{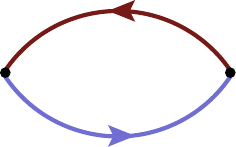}}}
\newcommand{\kaonnlo}{\raisebox{-0.4\totalheight}{\includegraphics[scale=.4]{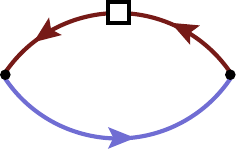}}}

\definecolor{codegray}{gray}{0.9}
\newcommand{\code}[1]{\colorbox{codegray}{\texttt{#1}}}

\begin{document}
\maketitle

\section{Introduction}

The computation of derivatives with respect to parameters of the action is a
tool that has proven to be very useful in lattice QCD simulations. Among others,
it allows to compute strong and electromagnetic isospin breaking effects---which
contribute at the percent level to hadronic observables like hadron mass
splittings and the HVP contribution to $g-2$---and also to correct for
mistunings in the choice of the bare parameters in lattice simulations.

For the case of strong isospin breaking, the different methods to introduce its
effects in lattice QCD simulations can be classified into two main groups:
\begin{itemize}
  \item Exact methods, in which the exact isospin breaking contribution is
        included for both dynamic and valence quarks. This category includes,
        for example, performing explicit simulations with non-degenerate up and
        down quark masses or performing exact reweighting in an already existing
        isospin-symmetric ensemble.
  \item Approximate methods, in which isospin breaking is introduced as a Taylor
        expansion in powers of the up-down quark mass splitting $\Delta m$
        truncated to some power. This category includes the RM123
        method~\cite{deDivitiis:2011eh}, or performing measurements at different
        quark masses in order to extract derivatives by, for example, taking
        finite differences.
\end{itemize}

The focus of the present study regards the inclusion of strong isospin breaking
effects using automatic differentiation through reweighting, which in essence
will be a generalization and full automation of the RM123 method up to arbitrary
orders.

This manuscript is organized as follows: in Sec.~\ref{sec:rm123}, we review the
RM123 method to include strong isospin breaking effects, specializing for the
case of the kaon correlation function; in
Sec.~\ref{sec:generalizing-rm123}, we explain how to generalize the RM123 method
to arbitrary orders in $\Delta m$ by introducing truncated polynomials along
with reweighting; in Sec.~\ref{sec:setup-results}, we explain the lattice setup
of our simulation and show our results comparing both approaches; finally, we
write our conclusions in Sec.~\ref{sec:conclusions-outlook}.

\section{Including isospin breaking effects with the RM123 method}
\label{sec:rm123}

Let us consider that we are interested in evaluating the expectation value
\begin{align}
  \label{eq:intro:target-integral}
\langle O \rangle_{\Delta m} = \frac{\int \mathcal{D}U\mathcal{D}\psi\mathcal{D}\bar{\psi}\; O\;
e^{-S_0 - S_{\mathrm{IB}}(\Delta m)}}{\int
\mathcal{D}U\mathcal{D}\psi\mathcal{D}\bar{\psi} \; e^{-S_0 - S_{\mathrm{IB}}(\Delta m)}},
\end{align}
in a theory with $N_{f}=1+1$ non-degenerate
quarks $\psi =[u,d]$ with an action that can be split into isospin-symmetric and
isospin-broken parts,
\begin{align}
S_0 = S_{G} + \sum_{f=u,d}^{}\bar{\psi}_{f}D \psi_{f} + m_{l} \sum_{x}^{} [\bar{u}u + \bar{d}d](x), \quad S_{\mathrm{IB}}(\Delta m) = -\Delta m \sum_{x}^{} [\bar{u}u - \bar{d}d](x),
\end{align}
where $S_{G}$ is the gauge action, $D$ is the massless Dirac
operator, $m_{l} =(m_{u}+m_{d}) / 2$ is the average of
and $\Delta m =(m_{d}-m_{u}) / 2$ the splitting between the masses $m_{u}$
and $m_{d}$ of the quarks $u$ and $d$.

While simulating two degenerate quarks can be done efficiently in QCD with the
standard Hybrid Monte Carlo algorithm \cite{Duane:1987de}, performing a simulation with
non-degenerate quarks requires the use of more complex algorithms---such as the
rational Hybrid Monte Carlo algorithm \cite{Kennedy:1998cu,Clark:2003na}---which
are computationally more expensive. The authors of Ref.~\cite{deDivitiis:2011eh}
proposed as an alternative to approximate $\langle O \rangle_{\Delta m}$ by
considering its Taylor expansion in $\Delta m$,
\begin{align}
  \label{eq:intro:taylor-expansion}
\langle O \rangle_{\Delta m} = \langle O \rangle_{0} +
\frac{\partial \langle O \rangle_{\Delta m}}{\partial \Delta m}\Big|_{\Delta m=0}\Delta m +
\mathcal{O}(\Delta m^2),
\end{align}
where $\langle O \rangle_{0}$ is an expectation value computed in the isospin-symmetric theory,
\begin{align}
\langle O \rangle_{0} = \frac{\int \mathcal{D}U \mathcal{D}\psi\mathcal{D}\bar{\psi}\; O\; e^{-S_{0}}}{\int
\mathcal{D}U\mathcal{D}\psi\mathcal{D}\bar{\psi} \; e^{-S_0}}.
\end{align}
If one knows the
derivative
$\partial \langle O \rangle_{\Delta m} / \partial \Delta m \big|_{\Delta m=0}$ and $\Delta m$ is small enough, one can
approximate $\langle O \rangle_{\Delta m}$ by truncating at first order
in $\Delta m$. Such derivative can be obtained by Taylor expanding the path
integral in Eq.~(\ref{eq:intro:target-integral}),
\begin{align}
  \label{eq:rm123:path-integral-expansion}
\langle O \rangle_{\Delta m} =& \frac{\int \mathcal{D}U\mathcal{D}\psi\mathcal{D}\bar{\psi}\; O(1+ \Delta m S_{\mathrm{IB}})
e^{-S_{0}}}{\int \mathcal{D}U\mathcal{D}\psi\mathcal{D}\bar{\psi}\; (1+ \Delta m S_{\mathrm{IB}}) e^{-S_{0}}} +
\mathcal{O}(\Delta m^2) = \frac{\langle O \rangle_{0} + \Delta
m\langle O S_{\mathrm{IB}} \rangle_{0}}{1 + \Delta m\langle S_{\mathrm{IB}}
\rangle_{0}} + \mathcal{O}(\Delta m^2) \nonumber \\
=& \langle O \rangle_{0} + \Delta m \langle O S_{\mathrm{IB}} \rangle_{0} + \mathcal{O}(\Delta m^2),
\end{align}
where we have used that $\langle S_{\text{IB}} \rangle_{0} = 0$ in the
isospin-symmetric theory. By comparing with
Eq.~(\ref{eq:intro:taylor-expansion}), we identify
\begin{align}
\frac{\partial \langle O \rangle_{\Delta m}}{\partial \Delta m}\Big|_{\Delta m=0} = \langle O S_{\text{IB}} \rangle_{0},
\end{align}
which can be evaluated in the cheaper, isospin-symmetric simulation.



\begin{figure}
  \centering
  \includegraphics[width=0.8\textwidth]{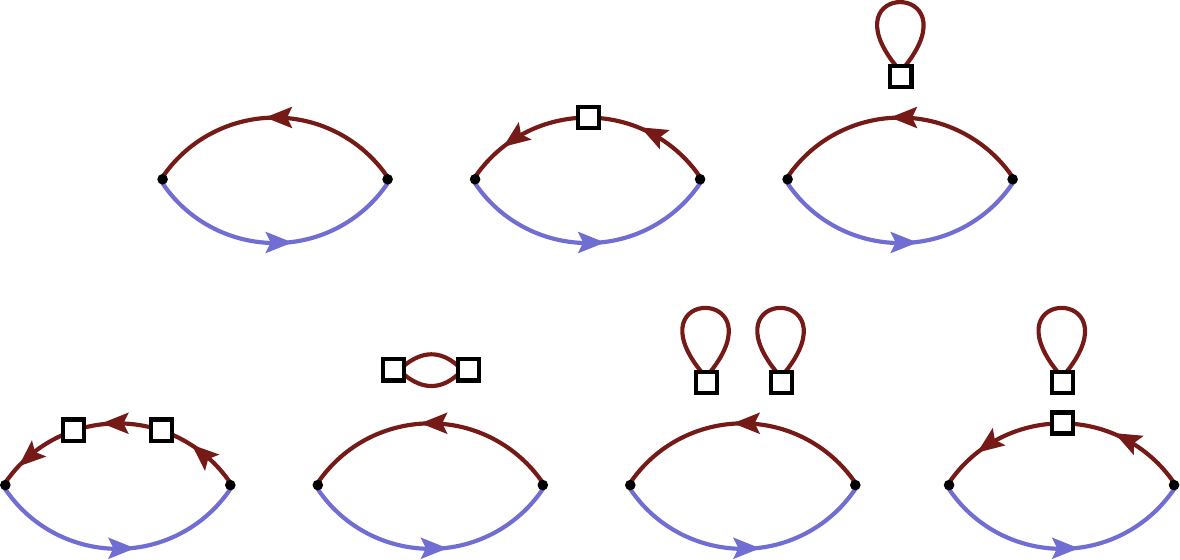}
  \caption{Diagrams resulting from evaluating the Wick contractions in
    Eq.~(\ref{eq:rm123:path-integral-expansion}) for the kaon
    correlation function in Eq.~(\ref{eq:rm123:kaon-corr-func}) to leading (top left), first
    order (top middle and top right), and second order (bottom row) in $\Delta m$. Filled black dots denote $\gamma_{5}$
    insertions, empty squares denote insertions of $S_{\mathrm{IB}}$, and red
    and purple lines denote up and strange quark propagator, respectively. }
  \label{fig:expeffects}
\end{figure}

In practice, this means that we can compute derivatives with respect to action
parameters by computing additional diagrams. Specializing to the case of the
charged kaon correlation function,
\begin{align}
  \label{eq:rm123:kaon-corr-func}
C_{K^{+}}(t) = \langle \bar{s}(t)\gamma_{5}u(t)\; \bar{u}(0)\gamma_{5}s(0) \rangle,
\end{align}
the leading order leads to the diagram in the top left of
Fig.~\ref{fig:expeffects} and the first order to the two diagrams in top middle
and right.\footnote{ The top right diagram cancels out when summing the
  contributions coming from the up and down quark~\cite{deDivitiis:2011eh}.} The
number of diagrams to compute increases combinatorily with the truncation order
of the Taylor expansion: if we were to truncate the Taylor expansion at second
order, we would need to compute the four diagrams in the bottom of the figure,
which need to be individually derived, implemented and tested. In the following,
we will propose and test a framework, based on automatic differentiation, which
fully automates the computation of all these diagrams up to arbitrary order.

\section{Generalizing the RM123 method with truncated polynomials}
\label{sec:generalizing-rm123}

\subsection{Truncated polynomials}

Truncated polynomials are a particular flavor of forward-mode automatic
differentiation~\cite{haro_ad}, and they rely on the fact that a differentiable
function $f(x)$ can be expanded in a Taylor series around $x^{(0)}$. Particularly,
\begin{align}
f(x^{(0)} + \Delta m) = f(x^{(0)}) + f'(x^{(0)})\Delta m + \frac{1}{2}f''(x^{(0)})\Delta m^2 + \dots
\end{align}
The construction of such Taylor expansion can be automated for any analytic function $f$ by using the algebra of truncated
polynomials:
if $\tilde{x} = \tilde{x}^{(0)} + \tilde{x}^{(1)}\Delta m + \dots + \tilde{x}^{(K)} \Delta m^{K}$ is a
truncated polynomial of order $K$ and we code all elementary mathematical
functions acting on these polynomials, and we evaluate $f(\tilde{x})$
with $\tilde{x} = \tilde{x}^{(0)} + \Delta m$, the output of the function will
be a truncated polynomial containing its first $K$ derivatives
\begin{align}
  f^{(k)}(\tilde{x}^{(0)}) = \frac{1}{k} \frac{\partial ^{k} f(x)}{\partial x^{k}} \Big|_{x = \tilde{x}^{(0)}}.
\end{align}
This machinery can be used, in principle, for any arbitrarily complex
function $f$, such as a computer program. Particularly, it has been succesfully
applied in lattice QCD for automatic error propagation in the analysis of Monte
Carlo data~\cite{Ramos:2018vgu}, as well as in lattice field theory simulations
to propagate derivatives with respect to action parameters throughout the HMC
algorithm or via reweighting techniques~\cite{Catumba:2023ulz}, to tackle the
sign problem in the quantum rotor~\cite{Albandea:2024fui}, and to tackle the
signal to noise problem in a $\phi^{4}$ theory~\cite{Catumba:2025ljd}.

In the present study we want to test if automatic differentiation can
successfully propagate derivatives when solving the Dirac equation, which is key
for the computation of matrix elements and hadron masses from correlation
functions such as the one in Eq.~(\ref{eq:rm123:kaon-corr-func}). As in the
previously mentioned applications, we will use the algebra of truncated
polynomials as implemented in \code{FormalSeries.jl}
\cite{alberto_ramos_2023_7970278} in the Julia programming language, and focus on
the computation of observables via reweighting.

\subsection{Reweighting with truncated polynomials}

As mentioned in the introduction, a possible way to introduce isospin breaking
effects in an observable is by reweighting an ensemble from $\Delta m = 0$ to a
particular $\Delta m \neq 0$~\cite{Finkenrath:2013soa,Finkenrath:2015ava}. To do
so, we write
\begin{align}
  \label{eq:rm123:reweighting-identity}
\langle O \rangle_{\Delta m} &= \frac{\int \mathcal{D}U\mathcal{D}\psi\mathcal{D}\bar{\psi}\; O\;
e^{-S_0 - S_{\mathrm{IB}}(\Delta m)}}{\int
  \mathcal{D}U\mathcal{D}\psi\mathcal{D}\bar{\psi} \; e^{-S_0 - S_{\mathrm{IB}}(\Delta m)}} = \frac{\int \mathcal{D}U\; O(\Delta m) W_{\mathrm{IB}}(\Delta m) \; e^{-S_{G}}}{\int \mathcal{D}U\; W_{\mathrm{IB}}(\Delta m)\; e^{-S_{G}}} \nonumber\\
  &= \frac{\langle O(\Delta m) W_{\mathrm{IB}}(\Delta m) \rangle_{0}}{\langle W_{\mathrm{IB}}(\Delta m) \rangle_{0}},
\end{align}
where in the second equality we have performed the integration over fermionic
fields and we have defined the isospin-breaking reweighting factor
\begin{align}
  \label{eq:rm123:reweighting-factor}
W_{\mathrm{IB}}(\Delta m) \equiv \frac{\det D(m_{l} - \Delta m) \det D(m_{l} + \Delta m)}{\det D^2(m_{l})},
\end{align}
and expectation values on the right-hand side are evaluated in the
isospin-symmetric theory. If we replace $\Delta m$ by the truncated polynomial
\begin{align}
  \label{eq:rm123:trun-pol-dm}
  \Delta \tilde{m} = \sum_{k=0}^{K} \Delta \tilde{m}^{(k)} \Delta m, \quad \Delta \tilde{m}^{(k)} =
\begin{cases}
  1 \quad &\text{for } k = 1 \\
  0 \quad &\text{for } k \neq 1
\end{cases},
\end{align}
evaluating Eq.~(\ref{eq:rm123:reweighting-identity}) will return the full
analytical dependence of the polynomial expansion of the observable $O$ up to
order $K$ centered around $\Delta m = 0$,
\begin{align}
\langle O \rangle_{\Delta m}^{(k)} = \frac{1}{k!} \frac{\partial ^{k}}{\partial (\Delta m)^{k}} \langle O \rangle_{\Delta m} \Big|_{\Delta m = 0}.
\end{align}

The evaluation of Eq.~(\ref{eq:rm123:reweighting-identity}) with truncated
polynomials requires, on the one hand, the computation of $O(\Delta \tilde{ m})$, which accounts
for the valence contribution of isospin-breaking effects. In our study we will
focus on the correlation function of the kaon, which, after integrating out
fermionic degrees of freedom, reads
\begin{align}
  \label{eq:rm123:kaon-corr-expansion}
  O(\Delta m) = \mathrm{tr} \left[ \gamma_{5}D^{-1}(m_{l}-\Delta m) \gamma_{5}D^{-1}(m_{s}) \right] = -\; \kaonlo - \kaonnlo \;\Delta m + \mathcal{O}(\Delta m^2),
\end{align}
where the dependence of the Dirac operator on the quark
masses is displayed for clarity and we have explicitly shown the diagramatic expansion of the
observable up to first order in $\Delta m$. While in the RM123 method one
needs to explicitly compute both diagrams by implementing
\begin{align}
  \kaonlo =& \;\mathrm{tr} \left[ \gamma_{5} D^{-1}(m_{l}) \gamma_{5}D^{-1}(m_{s}) \right], \\ \kaonnlo =& \; \mathrm{tr} \left[ \gamma_{5} D^{-1}(m_{l}) D^{-1}(m_{l}) \gamma_{5} D^{-1}(m_s) \right], \label{eq:rm123:o1}
\end{align}
the use of truncated polynomials allows to obtain the full polynomial expansion
of the observable by the evaluation of the left-hand side of
Eq.~(\ref{eq:rm123:kaon-corr-expansion}) with already existing code.

On the other hand, the computation of the reweighting
factor $W_{\mathrm{IB}}$ accounts for the
sea-quark contribution of isospin-breaking effects, and can also be evaluated
using the truncated polynomial in Eq.~(\ref{eq:rm123:trun-pol-dm}).
Since $\Delta \tilde{m}^{(0)} = 0$, the ratio in
Eq.~(\ref{eq:rm123:reweighting-factor}) is 1 at leading order and can be easily
estimated using $N$ complex normal stochastic sources $\eta_{i}$:
\begin{align}
W_{\mathrm{IB}} \equiv \det A = \int d \eta \; e^{-\eta^{\dagger} A^{-1} \eta} = \int d \eta \; e^{-\eta^{\dagger} \eta} e^{-\eta^{\dagger}(A^{-1} -1)\eta} \approx \frac{1}{N} \sum_{i=1}^{N} e^{-\eta_{i}^{\dagger}(A^{-1} -1)\eta_{i}},
\end{align}
where $A \equiv D_{u} D_{d} D^{-2}_{l}$. Note that, once estimated, $W_{\mathrm{IB}}$ can
be reused for any other observable $O(\Delta m)$.

In the following, we will focus on the computation of the valence contribution of
the kaon correlator in Eq.~(\ref{eq:rm123:kaon-corr-expansion}) with truncated
polynomials, where we will solve the Dirac equation by means of the conjugate
gradient algorithm.

\subsection{Conjugate gradient with truncated polynomials}

The conjugate gradient algorithm is an iterative algorithm that solves the Dirac equation
\begin{align}
D \psi = \eta.
\end{align}
Although truncated polynomials can be used with any computer program as long as
all mathematical operations acting on them are differentiable, it is not clear
that the propagated derivatives will converge as rapidly, nor if this is
guaranteed, for an iterative algorithm such as the conjugate gradient.  The main
objective of this study is to investigate whether truncated polynomials converge
to the right derivatives using the conjugate gradient algorithm by comparing
results with the RM123 method.

The only step which is non-differenciable in the conjugate gradient algorithm
and should be adapted to the use of automatic differentiation is the choice of
the stopping criterium. In the standard conjugate gradient, the usual choice is
to enforce that the residue of the solution is smaller than a particular
tolerance \code{tol}, i.e.
\begin{align}
  \label{eq:rm123:cg-lo-condition}
\left\| (D D^{-1} - I) \eta \right\|^2 < \mathrm{tol} \left\| \eta \right\|^2.
\end{align}
However, when using truncated polynomials what we will have is a Taylor
expansion in $\Delta m$ of our fermion fields,
\begin{align}
\eta \to \tilde{\eta} = \tilde{\eta}^{(0)} + \tilde{\eta}^{(1)} \Delta m + \tilde{\eta}^{(2)} \Delta m^2 + \mathcal{O}(\Delta m^3).
\end{align}
In particular, the residue of the solution will also become a truncated
polynomial: expanding $D D^{-1} - I$ in $\Delta m$, we have
\begin{align}
D D^{-1} - I - \Delta m \left[ D D^{-2}  - D^{-1}\right ] + \Delta m^2 \left[D D^{-3} - D^{-2} \right] + \mathcal{O}(m^3).
\end{align}
By analogy with the condition for the leading order in
Eq.~(\ref{eq:rm123:cg-lo-condition}), we decided to enforce that the residue
is smaller than a certain tolerance \code{tol} order by order,\footnote{Note that
  this stopping criterium is not the same as imposing
$\left[\left\| (DD^{-1}-I)\eta \right\|^{2}\right]^{(n)} < \mathrm{tol} \left\| \eta \right\|^2$ for
every order. Such an expansion up to order $\Delta m$ would be
\begin{align}
([DD^{-1}-I]\eta, [DD^{-1}-I]\eta) - \Delta m  \left\{([DD^{-2}-D^{-1}]\eta,[DD^{-1}-I]\eta) + ([DD^{-1}-I]\eta,[DD^{-2}-D^{-1}]\eta) \right\} + \mathcal{O}(\Delta m^2) \nonumber
\end{align}
which mixes contributions from different orders of the expansion of $DD^{-1}-I$.}
\begin{align}
  \label{eq:cg:condition-truncated-pol}
\left\| D D^{-n-1} \eta - D^{-n}\eta \right\| < \mathrm{tol} \left\| \eta \right\|^2.
\end{align}

\section{Setup and results}
\label{sec:setup-results}

In the following, we will investigate whether the condition in
Eq.~(\ref{eq:cg:condition-truncated-pol}) leads to a convergence of the first
order in $\Delta m$ when using truncated polynomials with the conjugate gradient
algorithm. For this proof-of-concept study we used a single ensemble,
generated using \code{openQCD}~\cite{Luscher2019_openQCD} as part of the
Coodinated Lattice Simulations (CLS) initiative~\cite{Bruno:2016plf}. It
has $N_{f}=2+1$ flavors of $O(a)$-improved Wilson fermions and a tree-level
improved Lüscher--Weisz gauge action, and other parameters of its simulation are
displayed in Tab.~\ref{tab:a654}.

\begin{table}
\begin{center}
\begin{tabular}{lccccccc}
\hline
ID & \(\beta\) & BC & \(V\) & \(a\) & \(M_{\pi}\) & \(M_{K}\) & \(M_{\pi} L\)\\
\hline
A654 & 3.34 & periodic & \(48 \times 24^{3}\) & 0.097 fm & 338 MeV & 462 MeV & 4.0\\
\hline
\end{tabular}
\end{center}
\caption{Parameters of the simulation used in this study: the bare
  coupling $\beta = 6 / g_{0}^{2}$, the boundary condition, the volume in lattice units, the lattice
spacing $a$ in physical units, and the approximate pion and kaon masses.}
  \label{tab:a654}
\end{table}

We focused on the computation of the derivative of the kaon mass with respect to
the up-down mass splitting, $\partial M_{K^{+}} / \partial \Delta m$, for which
we evaluated Eq.~(\ref{eq:rm123:kaon-corr-expansion}) using truncated
polynomials. For this purpose, we used a code
designed to perform lattice simulations on GPUs,
\code{LatticeGPU.jl}~\cite{Ramos2024_latticegpu,Catumba:2025jae}, written in
Julia, which has I/O compatibility with \code{openQCD}. Additionally, we used
the measurement code \code{GPUobs.jl}~\cite{Ramos2024_gpuobs}, which allows one
to obtain observables as truncated polynomials in $\Delta m$, and we set the solver
tolerance \code{tol} in Eq.~(\ref{eq:cg:condition-truncated-pol}) to
$10^{-28}$.

By evaluating Eq.~(\ref{eq:rm123:kaon-corr-expansion}) with truncated
polynomials of order 1, one obtains the power expansion
\begin{align}
  \tilde{C}_{K^{+}}(t) = C_{K^{+}}^{(0)} + C_{K^{+}}^{(1)} \Delta m + \mathcal{O}(\Delta m^2).
\end{align}
To extract the derivative of the kaon mass, we start by writing the large-time
behavior of the spectral decomposition of the kaon two-point function,
\begin{align}
  \label{eq:kaon-spectral-dec}
C_{K^{+}}(t) \propto A  \left[ e ^{-M_{K^{+}}(t - T / 2)} + e^{M_{K^{+}}(t - T / 2)} \right].
\end{align}
By expanding the matrix element and the kaon mass in powers of $\Delta m$,
\begin{align}
\tilde{A} = A^{(0)} + A^{(1)} \Delta m +  \mathcal{O}(\Delta m^2), \quad \tilde{M}_{K^{+}} = M_{K^{+}}^{(0)} + M_{K^{+}}^{(1)}\Delta m  + \mathcal{O}(\Delta m^2)
\end{align}
one can obtain $M_{K^{+}}^{(1)}$ by fitting to the relation
\begin{align}
  \label{eq:fit-m1}
\frac{C_{K^{+}}^{(1)}(t)}{C_{K^{+}}^{(0)}(t)} = \frac{A^{(1)}}{A^{(0)}} + M^{(1)}_{K^{+}}(t - T / 2) \tanh[M_{K^{+}}^{(0)}(t - T / 2)].
\end{align}

\begin{figure}
  \centering
  \includegraphics[width=0.49\textwidth]{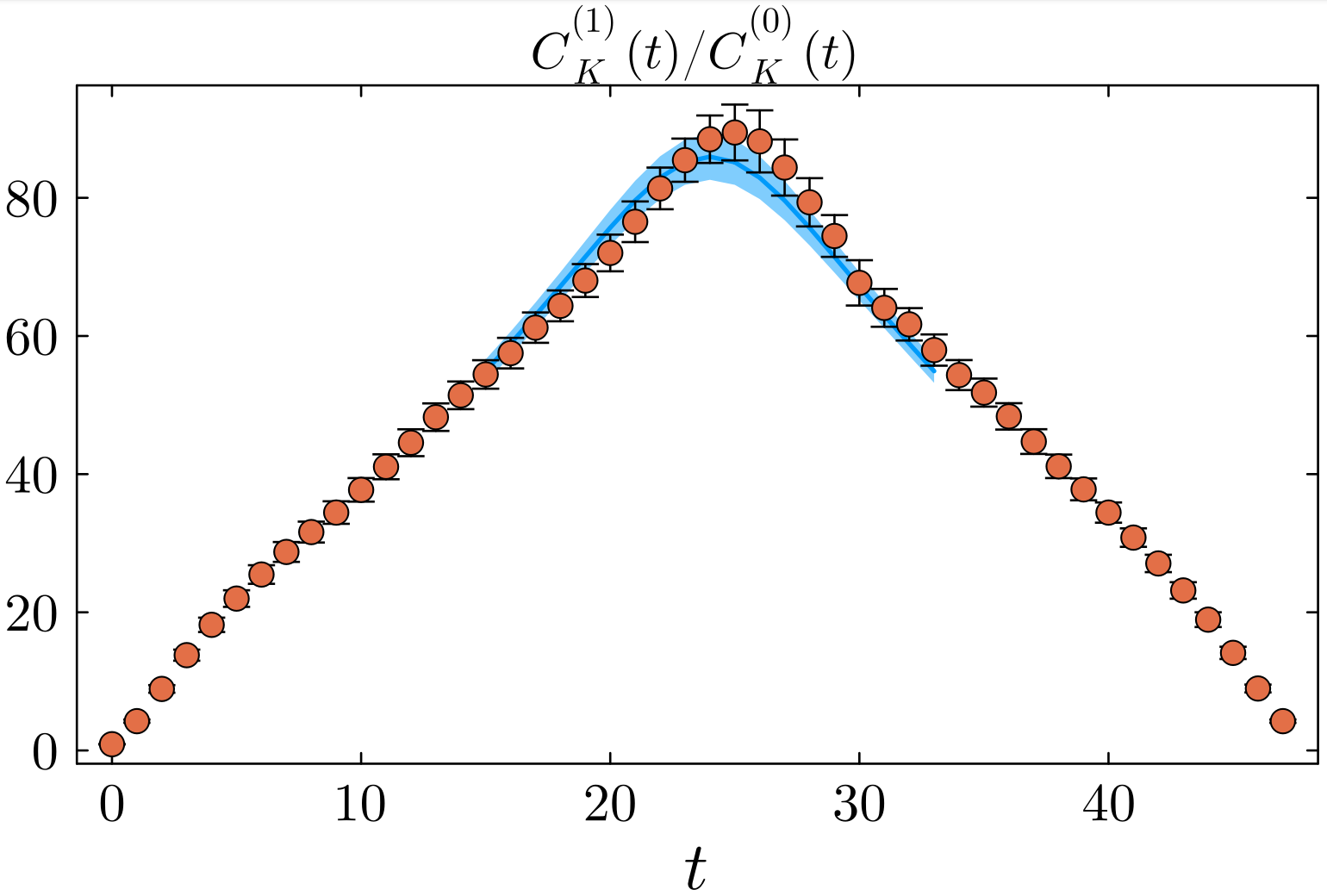}
  \includegraphics[width=0.49\textwidth]{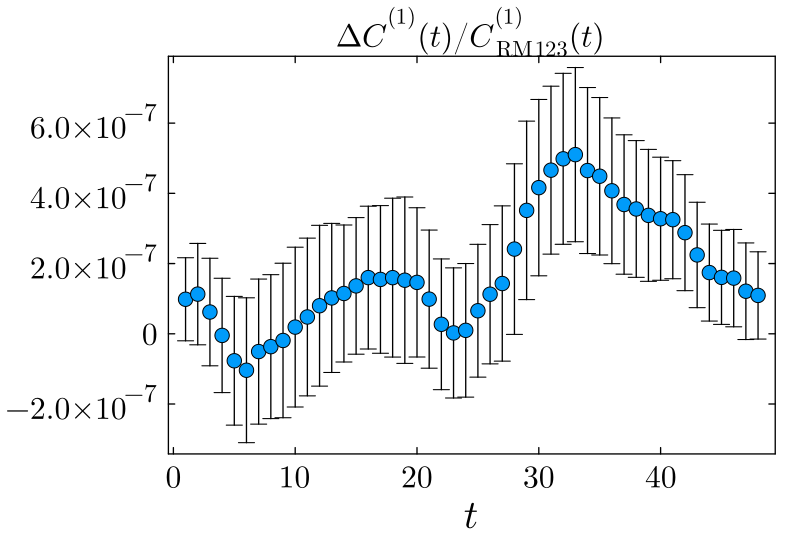}
  \caption{(Left) Ratio of first to leading order of correlation function of the
    kaon as a function of time. The blue band denotes the region fitted to the
    functional form in Eq.~(\ref{eq:fit-m1}). (Right) Relative difference
    between the first order of the correlation function of the kaon obtanied
    with truncated polynomials and RM123, and defined in Eq.~(\ref{eq:rm123-ad-diff}).}
  \label{fig:results}
\end{figure}

In Fig.~\ref{fig:results} (left), we show this ratio as a function of time, where the data was
obtained from a set of 100 configurations of the ensemble A654, and fitting to
the previous functional form we find
\begin{align}
  M_{K^{+}}^{(1)} = -3.86(52).
\end{align}
As a cross-check of this method, in Fig.~\ref{fig:results} (right) we plot the relative
difference between the derivative $C_{K^{+}}^{(1)}(t)$ obtained with automatic
differentiation and the one obtained with the RM123,
\begin{align}
  \label{eq:rm123-ad-diff}
\frac{\Delta C^{(1)}(t)}{C^{(1)}_{\mathrm{RM123}}(t)}\equiv \frac{C_{K^{+},\mathrm{RM123}}^{(1)}(t) - C_{K^{+},\mathrm{AD}}^{(1)}(t)}{C_{K^{+},\mathrm{RM123}}^{(1)}(t)}
\end{align}
where $C_{K^{+1},\mathrm{RM123}}^{(1)}$ is obtained by evaluating
Eq.~(\ref{eq:rm123:o1}). The fact that this relative difference is of the order
of $10^{-7}$ is an indication that derivatives with automatic
differentiation can be correctly propagated through the conjugate gradient
algorithm, and that the RM123 method up to arbitrary order can be fully
automated with truncated polynomials.

\section{Conclusions and outlook}
\label{sec:conclusions-outlook}

We have developed a new method to compute derivatives with respect to any action
parameter up to arbitrary order using truncated polynomials. Using this method,
we have studied the valence contribution of strong isospin breaking effects of
the kaon correlation function at first order in $\Delta m$. As a first proof of
concept, we have found that the first derivative of the kaon correlation
function computed with truncated polynomials converges and is compatible with
the one obtained with the RM123. A more thorough assessment of the solution's
precision and a comparison of the computational cost with respect to the RM123
method is currently a work in progress.

In the present study, we have restricted ourselves to the valence contribution
at first order in $\Delta m$. A natural next step is to extend the analysis to
higher orders in the mass expansion and to incorporate sea-quark contributions
by computing the reweighting factor in Eq.~(\ref{eq:rm123:reweighting-factor}).
Also, although we have focused on expansions in $\Delta m$ to describe strong
isospin-breaking effects, the framework is fully general and can be applied to
any action parameter, and could be used, among others, for the correction of
mistunings in the quark masses and for the inclusion of electromagnetic
isospin-breaking effects.

\section*{Acknowledgements}

We like to thank Alberto Ramos, Dalibor Djukanovic, Harvey B. Meyer, Hartmut
Wittig and Georg von Hippel for valuable discussions and for their comments on
this manuscript. We are grateful to our colleagues in the CLS initiative for
sharing ensembles.  Calculations for this project have been performed on the HPC
cluster Mogon-NHR at Johannes Gutenberg-Universit\"at (JGU) Mainz. The
estimation of statistical uncertainties of gauge averages in this work has been
performed using the $\Gamma$-method implemented in the \code{ADerrors.jl}
package~\cite{Ramos:2018vgu}. The diagrams corresponding to the Wick
contractions studied in this work were checked with the Mathematica package
\code{Quark Contraction Tool}~\cite{Djukanovic:2016spv}.

\bibliography{biblio.bib}
\addcontentsline{toc}{section}{References}
\bibliographystyle{JHEP}



\end{document}